\documentclass[aps,pre,twocolumn,superscriptaddress,showkeys]{revtex4-2}

\usepackage{graphicx}
\usepackage{xcolor}
\usepackage{hyperref}
\usepackage{amsfonts}
\usepackage{amsmath}
\usepackage[section]{placeins} % Force figs to stay in their section
\def\be{\begin{equation}}
\def\ee{\end{equation}}
\def\bea{\begin{eqnarray}}
\def\eea{\end{eqnarray}}

\newcommand{\rr}[1]{#1}
\newcommand{\reva}[1]{#1}

\begin{document}

%\preprint{}

\title{Enzyme regulation and mutation in a model serial dilution ecosystem}

\author{Amir Erez}
\thanks{These authors contributed equally}
\affiliation{Department of Molecular Biology, Princeton University, Princeton, New Jersey 08544, USA}
\affiliation{Racah Institute of Physics, The Hebrew University, Jerusalem, Israel}
\author{Jaime G. Lopez} 
\thanks{These authors contributed equally}
\affiliation{Lewis-Sigler Institute for Integrative Genomics, Princeton University, Princeton, New Jersey 08544, USA}
\author{Yigal Meir}
\affiliation{Department of Physics, Ben Gurion University of the Negev, Beer Sheva, Israel}
\author{Ned S. Wingreen}
\thanks{Corresponding author, email: wingreen@princeton.edu}
\affiliation{Department of Molecular Biology, Princeton University, Princeton, New Jersey 08544, USA}
\affiliation{Lewis-Sigler Institute for Integrative Genomics, Princeton University, Princeton, New Jersey 08544, USA}


\date{\today}

\begin{abstract}
Microbial communities are ubiquitous in nature and come in a multitude of forms, ranging from communities dominated by a handful of species to communities containing a wide variety of metabolically distinct organisms. This huge range in diversity is not a curiosity - microbial diversity has been linked to outcomes of substantial ecological and medical importance. However, the mechanisms underlying microbial diversity are still under debate, as simple mathematical models only permit as many species to coexist as there are resources. A plethora of mechanisms have been proposed to explain the origins of microbial diversity, but many of these analyses omit a key property of real microbial ecosystems: the propensity of the microbes themselves to change their growth properties within and across generations. In order to explore the impact of this key property on microbial diversity, we expand upon a recently developed model of microbial diversity in fluctuating environments. We implement changes in growth strategy in two distinct ways. First, we consider the regulation of a cell's enzyme levels within short, ecological times, and second we consider evolutionary changes driven by mutations across generations. Interestingly, we find that these two types of microbial responses to the environment can have drastically different outcomes. \rr{Enzyme regulation may collapse diversity over long enough times while, conversely, strategy-randomizing mutations can} produce a ``rich-get-poorer'' effect that promotes diversity. This work makes explicit, using a simple serial-dilutions framework, the conflicting ways that microbial adaptation and evolution can affect community diversity.
\end{abstract}

%\keywords{}

\maketitle

\section*{Introduction}

Microbial communities are a key component of nearly every ecosystem, ranging from arctic sediments \cite{Bienhold2012} to the human digestive tract \cite{Smits2017}. The composition of these communities can vary dramatically, ranging from communities dominated by a small number of metabolically similar organisms \cite{gajer2012temporal}, to communities composed of hundreds of metabolically diverse organisms \cite{Lloyd-Price2016,Ladau2013,Weigel2019}. Even within a given type of ecosystem there can exist substantial variation in community form \cite{gajer2012temporal}. This huge variation in microbial diversity is not merely a theoretical curiosity, having been linked to outcomes ranging from ecosystem stability to the results of medical treatments \cite{Ptacnik2008, VanElsas2012, Taur2014, Stein2013}. Thus, to better understand and engineer ecosystems, a strong theoretical understanding of the drivers of microbial diversity is required. 

Early theoretical work on ecological diversity led to the competitive exclusion principle, a prediction that the number of coexisting species in an ecosystem at steady state will not exceed the number of nutrients \cite{Levin1970, Armstrong1980}. However, it became apparent that many communities sustain diversity far in excess of what is predicted by competitive exclusion, famously exemplified by Hutchinson's ``paradox of the plankton'' \cite{Hutchinson1961}. This apparent clash between theory and observations has led to decades of study, attempting to bridge the gap and to develop an understanding of what drives diversity. Of the many important mechanisms for maintenance of diversity beyond competitive exclusion, we mention a few: microbial interactions \cite{Goyal2018, Kelsic2015}, predation \cite{Thingstad2000,PhysRevLett.119.268101}, spatial heterogeneity \cite{Murrell2003, Tilman1994}, non-steady-state dynamics \cite{Hutchinson1961,scheffer2003plankton}, and resource competition with trade-offs \cite{Posfai2017,Weiner2019,ErezLopez2020,li2020modeling,PaccianiMori2020}. 

The majority of the theoretical work on microbial diversity has relied on a chemostat framework in which nutrients are continuously supplied \cite{Palmer1994}. Often, however, in both natural and experimental microbial ecosystems  nutrients are supplied at time intervals, instead of being constantly supplied. In natural ecosystems, this reflects the passage of seasons \cite{Smits2017} or other environmental fluctuations. In experimental ecosystems, this reflects the commonly used serial dilution protocol in which microbes are periodically diluted and supplied with a fresh bolus of nutrients \cite{Lenski, Goldford2018}. Thus, further theoretical work is needed to understand diversity in systems where nutrients are supplied in a non-constant manner.

Recently, we considered diversity in a serial dilution ecology consisting of microbes competing for multiple nutrients \cite{ErezLopez2020}. Each species was defined by a strategy vector which quantifies its ability to uptake different nutrients. In the framework we had proposed, each species had a fixed and unchangeable `enzyme budget' it allocated. Strikingly, we found that unlike steady-state ecosystems, diversity was strongly dependent on the amount of nutrient supplied to the community, and that the changes in diversity could be understood as arising from an `early-bird effect'. In this early-bird effect, a species whose strategy allows it consume the most easily available nutrients gains an early population advantage and is then able to outcompete competitors for less-available nutrients, even if the early-bird species is a less efficient consumer of the latter. This effect is generally strengthened with increasing nutrient supply, though in certain cases the effect can be eliminated by saturating concentrations of nutrients. As a result, the long-term community composition depends on the amount of nutrients supplied to the ecosystem. If the early-bird species is abundant at low nutrient supply, this effect leads to a decreasing community diversity with increasing nutrient supply, with the opposite occurring if the early-bird species is low-abundance at low nutrient levels. In more complex scenarios this effect can lead to non-monotonic relationships between diversity and nutrient supply. 

In our previous investigations of serial-dilution models, the metabolic strategy of each species was unchanging throughout time. However, in reality, microbes can and do change their nutrient uptake strategies over both ecological and evolutionary timescales. On ecological timescales, bacteria have the ability to regulate their enzyme production, thus responding to environmental changes by shifting their strategy \cite{shimizu2014regulation,bajic2020ecology}. On evolutionary timescales, random mutations can lead to hardwired changes in the strategies of bacterial species \cite{Good2018,ramiro2020low,friedman2013sympatric}. How might such changes in metabolic strategies impact microbial diversity? 

In this work, we probe how both adaptation through enzyme regulation and mutation influence diversity in a serial dilution ecosystem. We find that these two forms of response to environmental pressures produce substantially different results. Mutations increase diversity relative to a model with unchanging metabolic strategies, particularly so when there is a large amount of growth between dilutions. In contrast, we find that diversity in a stable community can be curtailed by the addition of a species capable of sensing ambient nutrient concentration and thereby regulating its enzyme strategy. Interestingly, the destruction of diversity by such an \emph{adapter} species occurs on an emergent timescale, which can be much longer than any intrinsic timescale directly appearing in the dynamics. 

\section*{Results}
\begin{table}
	\caption{Annotation glossary}
	\centering
	\resizebox{\columnwidth}{!}{
	\begin{tabular}{ll}
		\hline 
		\emph{Symbol}			& \emph{Description} \\
		\hline
		$t$						& Time measured from the beginning of a batch \\
		$p$						& Number of nutrients\\
		%$\vec{x}$		    		& All vectors are $p$-dimensional \\
		$m$						& Number of competing species \\
		$m_e$					& Effective number of species at steady state \\
		$\nu$					& Mutation rate \\
		\hline
		$i$						& $(1...p)$ Latin index enumerating nutrients \\
		$c_i(t)$ 				& Time dependent concentration of nutrient $i$ \\
		$c_0$					& $\sum_{i=1}^{p} c_i(0)$; total nutrient concentration at time $t=0$ \\
		$K_i \equiv K$			& Monod half-velocity constant \\
%		$I_i$					& $\int_{0}^{\infty} \frac{c_i}{K_i+c_i}\,dt$; nutrient Monod function time integral \\
		\hline
		$\sigma,\sigma',...$		& $(1...m)$ Greek indices enumerating species \\
		$\rho_\sigma(t)$			& Species $\sigma$ biomass density at time $t$ since the start of a batch \\
		$x_\sigma(t)$			& Species $\sigma$ relative abundance at time $t$ since the start of a batch \\
		$\vec{\alpha}_\sigma$ 	& $(\alpha_{\sigma,1},..., \alpha_{\sigma,p})$; enzyme allocation strategy for species $\sigma$ \\
%		$\varepsilon$			& Standard deviation in enzyme budget \\
		$E$					& $E=\sum_i \alpha_{\sigma, i} = 1$ ; enzyme budget \\
		\hline
%		$\Gamma^\sigma_{i,i'}$		& Byproduct matrix converting nutrient $i'$ to nutrient $i$ \\
		$j_{\sigma,i}$			& Nutrient $i$ consumption rate by species $\sigma$ \\
%		$Y_{i}$			& Biomass yield on nutrient $i$\\
		\hline
	\end{tabular} }
\end{table}
The models that we explore in this paper are built on a generalized serial dilution framework in which $m$ species compete for $p$ nutrients within a series of recurring batches. Beginning each batch, a bolus of nutrients is provided such that $c_0 = \sum_i^p c_i(0)$, where $c_i(0)$ is the concentration of nutrient $i$ at time 0, measured from the beginning of the batch. At the same time 0, microbes are seeded into the batch in an inoculum of species such that $\rho_0 = \sum_{\sigma}^{m} \rho_\sigma(0)$, where $\rho_\sigma(t)$ is the biomass concentration of species $\sigma$ at time $t$ from the beginning of the batch. After all nutrients are exhausted within a batch, a new batch is initialized with the same nutrient bolus and an inoculum of total concentration $\rho_0$ whose composition is proportional to the species composition at the end of the previous batch. In short, we inoculate each batch with microbes and nutrient, wait for the microbes to consume the nutrient, and then dilute the resulting species composition to use as the inoculum for the next batch, and so forth.

A species, indexed by $\sigma$, is defined by its strategy vector $\vec{\alpha}_\sigma = (\alpha_{\sigma,1},...,\alpha_{\sigma,p})$, where $\alpha_{\sigma,i}$ is the maximum uptake rate of nutrient $i$ for species $\sigma$. The uptake rates are defined by Monod functions, $j_{\sigma,i} = \frac{c_i}{K_i+c_i}\, \alpha_{\sigma,i}$, where $j_{\sigma,i}$ is the uptake rate of nutrient $i$ by species $\sigma$ and $K_i$ is the half-saturation constant of nutrient $i$. For simplicity, we assume that $K_i \equiv K$ (we explored unequal $K_i$ in \cite{ErezLopez2020}). From the uptake rates, we can define the nutrient and population dynamics within a batch:

\bea 
\label{eq:dynamics}
\frac{dc_i}{dt} &=& -\sum_\sigma \rho_\sigma j_{\sigma,i}, \\
\label{eq:dynamicsrho}
\frac{d\rho_\sigma}{dt} &=& \rho_\sigma \sum_i j_{\sigma,i}.
\eea

\begin{figure}
	\includegraphics[width=0.99\columnwidth]{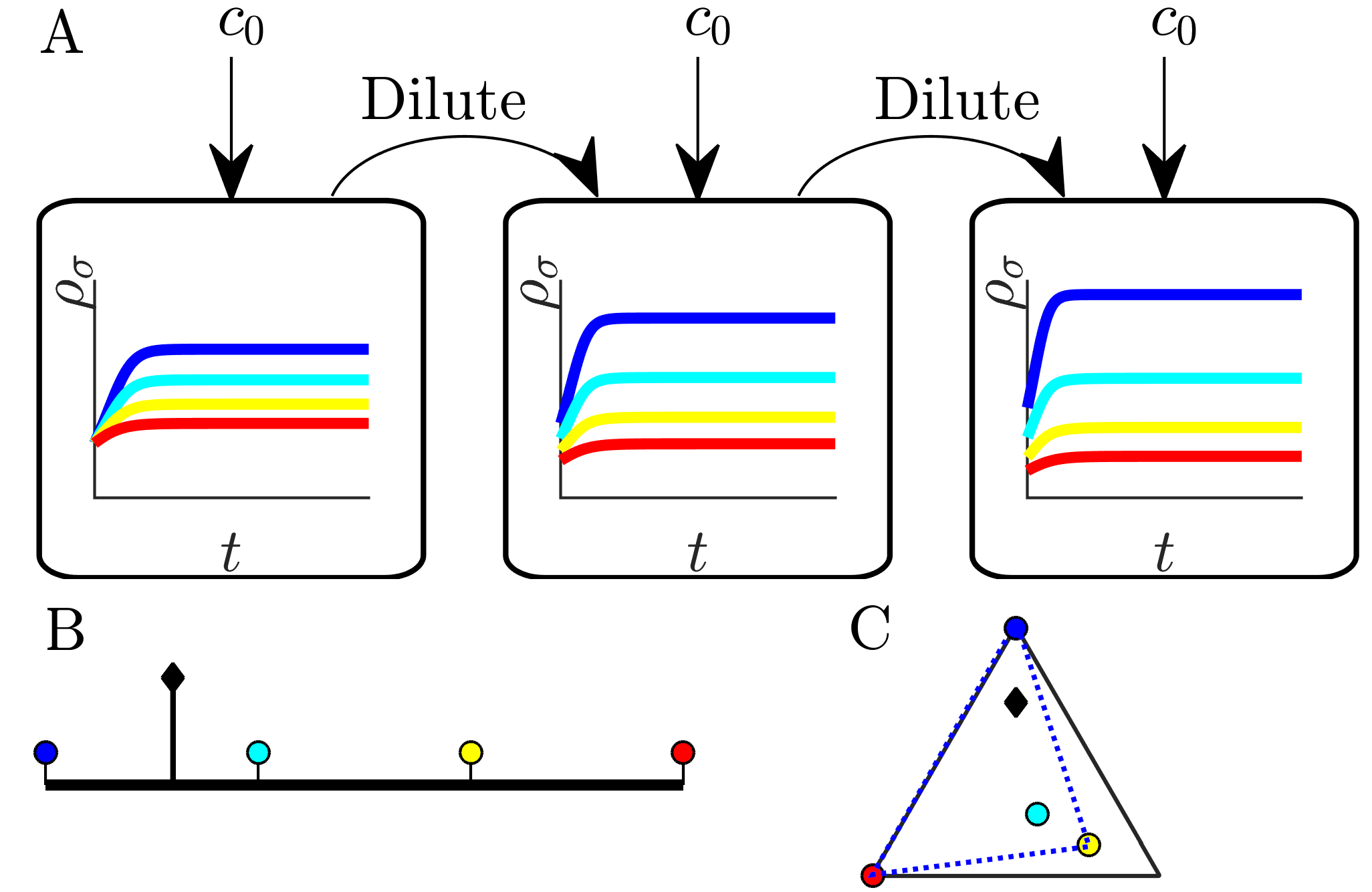}
	\caption{Reproduced from \cite{ErezLopez2020} under the  \href{https://creativecommons.org/licenses/by/4.0/}{Creative Commons Attribution}. Illustration of serial dilution resource-competition model. (\emph{A}) Serial dilution protocol. Each cycle of batch growth begins with a cellular biomass density $\rho_0$ and total nutrient concentration $c_0$. The system evolves according to Eqs.~\ref{eq:dynamics}-\ref{eq:dynamicsrho} until nutrients are completely consumed. A fraction of the total cellular biomass is then used to inoculate the next batch again at density $\rho_0$. (\emph{B}) Representation of particular enzyme-allocation strategies $\{\alpha_\sigma\}$ (colored circles) and nutrient supply composition $c_i/c_0$ (black diamond) on a 2-nutrient simplex, where the right endpoint corresponds to \reva{$c_1/c_0=1$}. (\emph{C}) Representation of particular strategies (circles) and nutrient supply (black diamond) on a 3-nutrient simplex. \reva{Dashed blue - the convex hull of the enzyme-allocation strategies. Here, the nutrient supply (black diamond) is inside the convex hull, implying coexistence of all species in the chemostat limit (see text).}}
	\label{fig:fig1}
\end{figure}

These dynamics are represented graphically in Fig. \ref{fig:fig1}A. The ``steady state'' of this deterministic ODE system is not a single fixed point, but instead an entire batch timecourse such that the relative species abundance at the beginning and end of the batch are identical. Explicitly, at steady state, whereas within a batch $\frac{d}{dt} \ne 0$, the inoculum, $\rho_\sigma(t=0)$, does not change from one batch to the next. 

Microorganisms typically operate near their biophysical limits \cite{Bialek10040}, capping their total protein-production capacity. Since capacity must be allocated for a multitude of essential cellular processes, microbes have a limited capacity to manufacture the enzymes used to consume nutrients. Roughly speaking, one would expect that microbes which are found to coexist would evolve similar metabolic enzyme production capacities. We take this viewpoint, thereby constraining our model to a fixed total amount of enzymes in the strategy vector: $E = \sum_i \alpha_{\sigma,i} = \mbox{const}$ for all species (relaxing this constraint leads to extinction of species with lower enzyme budgets and, depending on the other timescales in the system, determines the long-term diversity \cite{ErezLopez2020}). The fixed sum means that the strategies of different species can be represented on a simplex, shown as circles in Fig. \ref{fig:fig1}B and C, with the relative composition of the nutrient bolus represented as a black diamond.

In our earlier work \cite{ErezLopez2020}, we extensively characterized the behavior of this model when the metabolic strategies of the species are fixed over time. When the nutrient bolus is small ($c_0/K \ll 1$), we found that the serial dilution ecosystem behaves in a chemostat-like manner. In this limit, the system can support an unlimited number of coexisting species as long as the convex hull of the strategies (visualized in two dimensions as stretching a rubber band around the strategies) contains the nutrient composition. Examples of communities where this condition is met are shown in Fig. \ref{fig:fig1}B and C. We found that as more nutrient is provided to the system, i.e., the bolus size $c_0$ grows larger, this convex hull rule still applies but with `remapped' convex hull nodes. These remapped nodes generally differ from the original strategies and move as a function of bolus and inoculum size. This remapping can lead to large shifts in community diversity, with the direction of the shift determined by ecosystem details. 

\begin{figure}%[tbhp]
	\centering
	\includegraphics[width=\columnwidth]{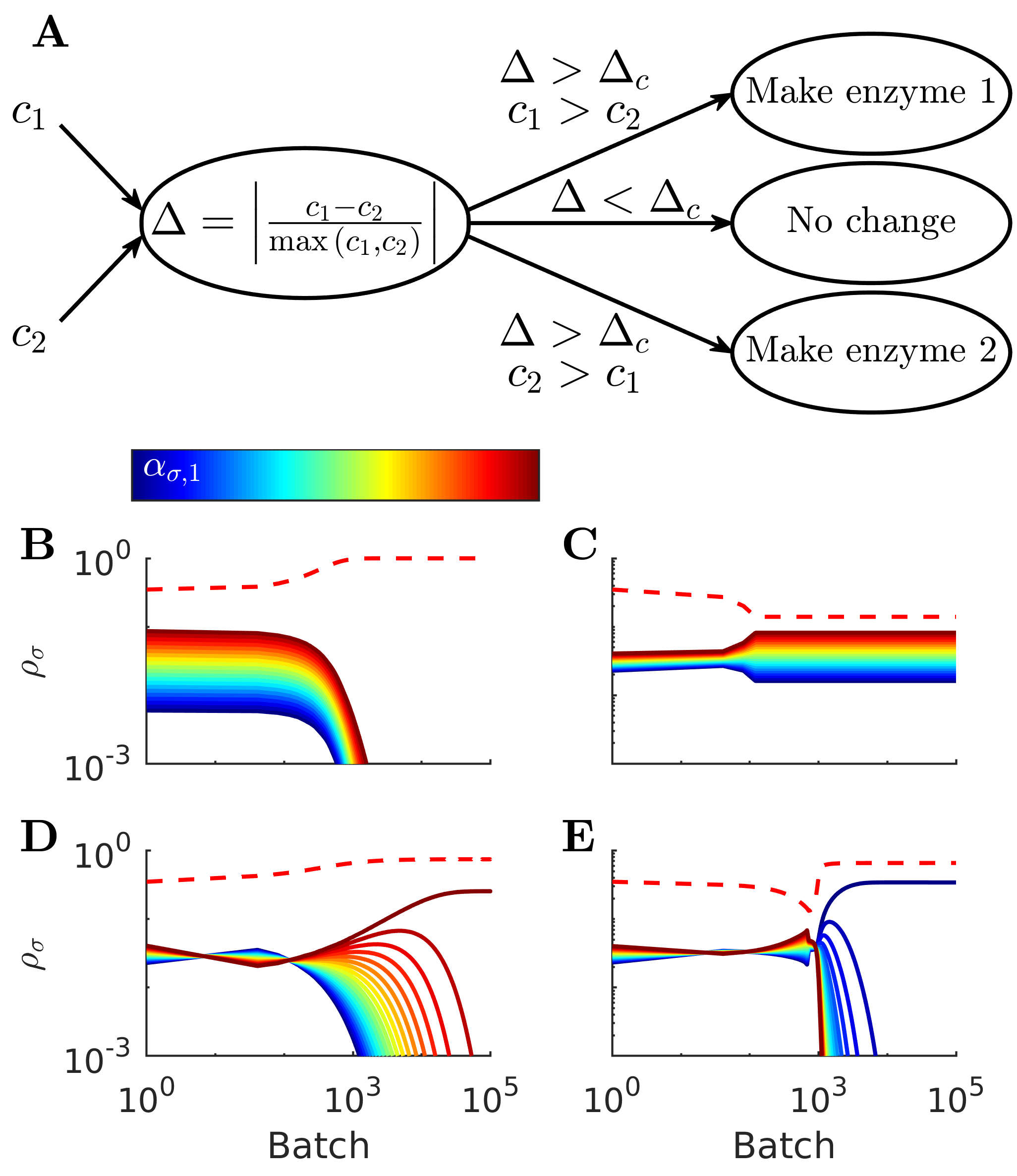}
	\caption{
		Serial dilution model with enzyme regulation. (\emph{A}) Schematic of adapter control scheme. The adapter changes which enzyme it produces in response to changing nutrient concentration. If the relative difference between the two nutrients is greater than the sensing threshold $\Delta_c$, it switches production to the enzyme corresponding to the more abundant nutrient. (\emph{B}-\emph{E}) Representative long-term dynamics of serial dilution communities after the addition of an adapter. The adapter population is shown by the dashed red curve. Communities containing 21 species with equally spaced strategies (cf. Fig \ref{fig:fig1}B) were allowed to reach steady state before the community was perturbed by an invasion that replaced 35\% of the community biomass with an adapter. (\emph{B}) Community growing with $\rho_0 = c_0 = 1$, and $\mbox{Nutrient 1 fraction} = 0.7$ invaded by an adapter with $\Delta_c = 0.02$. (\emph{C}) Community: $\rho_0 = 1$, $c_0 = 10^2$, and $\mbox{Nutrient 1 fraction} = 0.55$; adapter: $\Delta_c = 0.25$. (\emph{D}) Community: $\rho_0 = 1$, $c_0 = 10^2$, and $\mbox{Nutrient 1 fraction} = 0.55$; adapter: $\Delta_c = 0.02$. (\emph{E}) Community: $\rho_0 = 1$, $c_0 = 10^3$, and $\mbox{Nutrient 1 fraction} = 0.55$; adapter: $\Delta_c = 0.25$. All communities simulated with $K = 1$}
	\label{fig:fig6}
\end{figure}

%\subsection*{Enzyme regulation}
\emph{Enzyme regulation}. Bacteria are able to dynamically control the levels of their enzymes in response to changes in the environment \cite{shimizu2014regulation}. How would such regulation impact population dynamics in our model? To address this question, we introduce into our model an organism with the ability to reallocate its enzyme budget, and we call it the \emph{adapter}. The adapter's enzyme pool is continuously diluted by growth - typically in nature when a bacterial cell divides, approximately half the enzyme content goes to each daughter. We consider the enzymes as continuously replenished with newly produced enzymes, and the \emph{adapter} is able to select which type(s) of enzyme to produce. In general, for a given nutrient environment increasing one type of enzyme will yield the greatest increase in growth rate, so we consider the adapter to produce one type of enzyme at a time. As a result, the model equations are now generalized to allow the adapter's enzyme composition to have dynamics, according to $\frac{d}{dt} \alpha_{\sigma^*,i} = (\mathbb{P}_{\sigma^*,i} - \alpha_{\sigma^*,i} )\sum_{i'} j_{\sigma^*,i'}$, where $\sigma^*$ indexes the adapter species and $\mathbb{P}_{\sigma^*,i}$ is an indicator function that is unity when the adapter is producing enzyme $i$. 

We consider a two-nutrient case ($p=2$) in which the adapter switches its enzyme production to match the most abundant nutrient, as shown in Fig.~\ref{fig:fig6}\emph{A}. The adapter senses a relative difference, $\Delta$, between the two nutrients, defined as $\Delta = \left| \frac{c_1 - c_2}{\max{(c_1,c_2)}} \right|$. To model sensory uncertainty, the adapter can only respond to relative differences above a certain magnitude, $\Delta_c$. We assume that the time between batches is short enough that the adapter maintains both its enzymes levels and enzyme-production state between batches. This ability to sense the environment and modify its enzyme production allows the adapter to better exploit any early-bird advantages it gains. The adapter can modify its enzyme strategy to efficiently consume initially abundant nutrients and then further modify its strategy to consume the remaining nutrients. As a result, the adapter will have a fitness advantage over non-adapters if the changes in nutrient availability exceed its sensing tolerance. \rr{Thus, adaptation in our model can lead to diauxie - a widely observed phenomenon in which different nutrient types are consumed in sequence rather than simultaneously (see \textit{Appendix} Fig. \ref{fig:diauxie_example}).}

\indent\indent How does a community of non-adapters respond to sudden invasion by an adapter? To explore this, we allowed communities of 21 non-adapter species to reach steady state before replacing a fraction of the community biomass with an adapter. We tested a wide range of initial community and adapter parameters and show four different invasion outcomes in Fig.~\ref{fig:fig6}\emph{B-E}. Despite enzyme regulation occurring on very short timescales (on the scale of the adapter's doubling time), we find that the adapter introduces an emergent long timescale over which the community population changes. In the examples, this new timescale is on the order of $10^3$ batches, substantially longer than the $\sim$10 batches required for the initial communities to come to steady state. Moreover, this new timescale can be made even longer by starting with a smaller initial adapter biomass.

\indent\indent The post-invasion steady states of the communities can vary based on initial community and adapter parameters. In many cases, after being invaded the ecosystem gradually moves towards extinction of most of the non-adapter species, as is shown in Fig. ~\ref{fig:fig6}\emph{B,D,E}. However, in certain cases, such as that in Fig.~\ref{fig:fig6}\emph{C}, the adapter can coexist with the non-adapter community. This outcome occurs when the community becomes organized such that $\Delta<\Delta_c$ $\forall t$ within a batch. In other words, diversity is robust to adapter invasion if the community self-organizes to a state where the relative difference between the two nutrients never exceeds the adapter threshold tolerance. Under these conditions, the adapter loses the ability to switch its enzyme production and effectively becomes locked as a non-adapter specialist, i.e. consuming only one type of nutrient.. Therefore, the community of non-adapting species may lock the adapter into a fixed state thereby eliminating its inherent advantage over non-adapting species. In addition to reaching various final steady states, the dynamics of this system en route to steady state can also vary widely. For example, in Fig.~\ref{fig:fig6}\emph{B} there is a monotonic decrease in non-adapter abundances. In contrast, the non-adapter dynamics in Fig.~\ref{fig:fig6}\emph{C-E} are non-monotonic, with the abundance profile of the non-adapters being inverted multiple times in Fig.~\ref{fig:fig6}\emph{D-E}.

The adapter's slow takeover of communities indicates that enzyme regulation confers a small fitness advantage. This advantage may be offset by the cost of sensing and responding to environmental conditions - a cost which we do not model here. Moreover, as in the case with unequal enzyme budgets, the ecological relevance of the timescale introduced by the adapter will depend on other timescales in the system, such as that introduced by immigration. Furthermore, the effect of an adapter may be mitigated by noisy population dynamics: an invading adapter with a small population is sensitive to random extinctions, as the adapter's fitness is only slightly greater than that of the residents. In summary, though an adapter has a fitness advantage, it is not guaranteed that this advantage will translate to the adapter taking over the system.

\begin{figure}%[tbhp]
	\centering
	\includegraphics[width=\columnwidth]{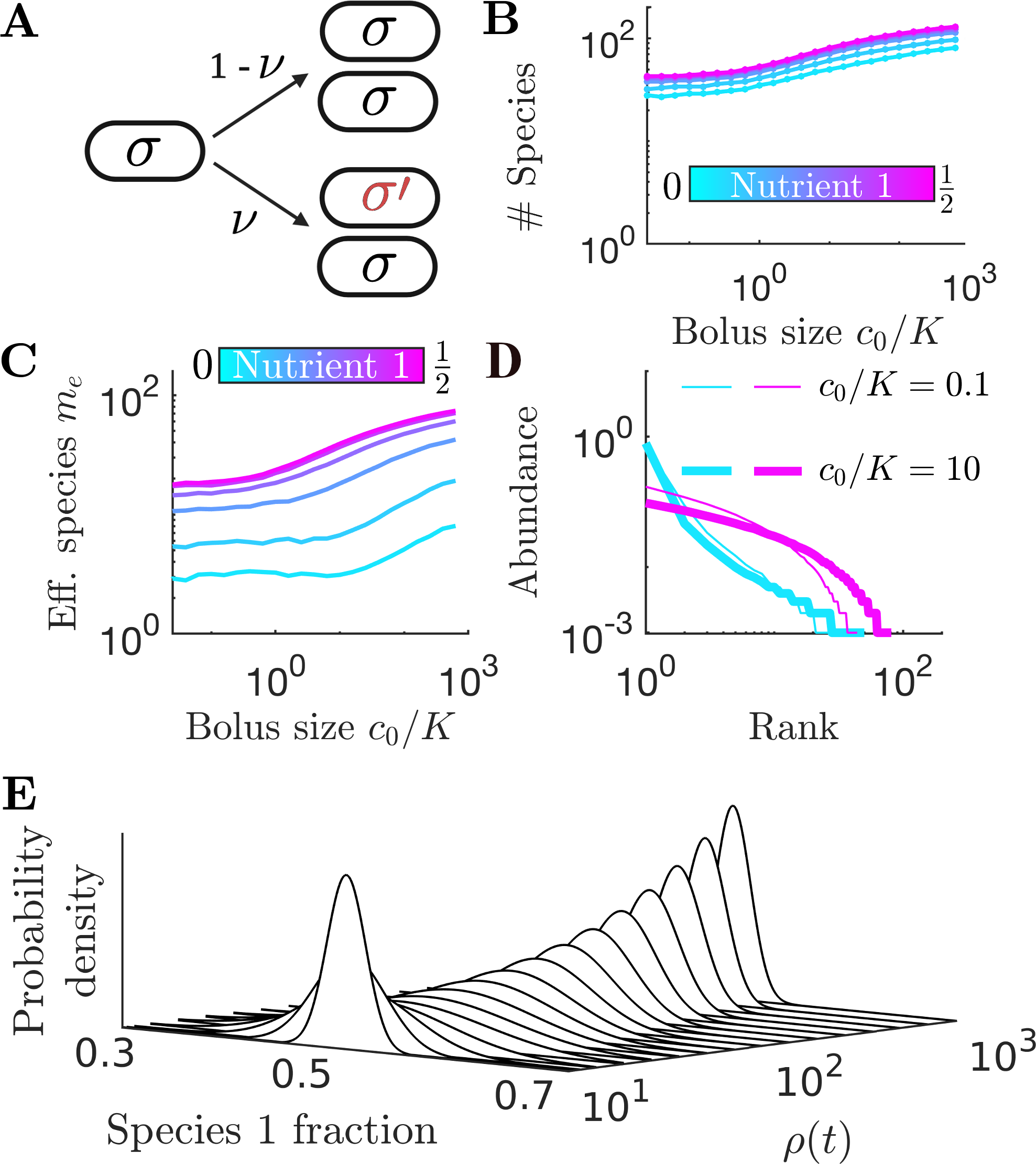}
	\caption{Diversity of species under mutation-selection balance. Starting from an inoculum of 1000 cells, with $K=1000$ and a fraction $\nu=0.01$ of cell divisions results in a mutation to a randomly-selected one of 201 evenly spaced strategies. Populations recorded at the start of each batch. (\emph{A}) Schematic of mutation. Each division either produces two daughter cells identical to the parent (with probability $1- \nu$) or one daughter cell identical to the parent and one daughter cell of a random strategy (with probability $\nu$). (\emph{B}) The median number of extant species under mutation-selection balance, versus $c_0/K$ for varying supply proportions, recorded at the start of each batch. (\emph{C}) Effective number of species $m_e$ for different nutrient compositions (colors) as a function of nutrient bolus size $c_0/K$. (\emph{D}) Rank-abundance curves for Nutrient 1 fractions 0.05 (cyan) and 0.5 (magenta); line thickness indicates value of $c_0/K$. \rr{ (\emph{E}) Numerical simulation of illustrative two-species Fokker-Planck model for the effect of mutations within a batch  (see Eq. \ref{eq:pde}). We initialize the simulation with a narrow distribution of abundances centered around species 1 fraction $x_1 = 0.5$ with $\rho_0 = 10$ and $\nu = 0.5$. } }
	\label{fig:mutate}
\end{figure}

%\subsection*{Mutation-Selection Balance}
%\indent \indent 
\emph{Mutation-Selection Balance}. To bring in one of the main drivers of diversity in the wild, we extend our model to include mutations and the resulting mutation-selection balance. Specifically, we introduce mutations as random changes in metabolic strategy \cite{Good2018}. Essentially, instead of allowing a single adapter species to modify its strategy, we let the repertoire of fixed strategies evolve and compete. Since a mutant is initially present as a single cell, it becomes essential to stochastically model the population dynamics, including both reproduction and sampling for each inoculum. Within a batch, instead of the deterministic ODEs of Eqs.~\ref{eq:dynamics}-\ref{eq:dynamicsrho} we simulate stochastic dynamics using Gillespie's method, summarized in Table~\ref{tab:Gillepsie} in the \emph{Appendix}. For large populations, the resulting steady state matches the deterministic one. To account for mutations, we modify the growth term to allow for \emph{mutation} events, whereby when species $\sigma$ increases by one cell, instead of making another $\sigma$, it sometimes makes a $\sigma'$ cell, i.e. $\sigma \rightarrow \sigma + \sigma'$. \emph{Mutation} occurs at a rate $\nu \rho_\sigma \sum_i j_{\sigma,i}$, while normal growth, $\sigma \rightarrow 2\sigma$, occurs at a rate $(1-\nu) \rho_\sigma \sum_i j_{\sigma,i}$. Fig. \ref{fig:mutate}A shows a schematic of this process. Together, stochastic reproduction, intra-batch mutations, and inter-batch sampling lead to complex dynamics whereby a species can appear, flourish for a number of batches, then die out, with different species replacing it. This results in fluctuations in the number of species present from batch to batch (\emph{Appendix} Fig.~\ref{fig:DilutionRegimes}). %In this manuscript we do not address heritable mutations, though heritability may influence the mutation-selection balance \cite{Good2018}. Nevertheless, the framework presented here may be readily extended to accommodate heritable mutations.

How does species diversity depend on nutrient bolus size $c_0$ under conditions of mutation-selection balance? As one would expect, the larger the nutrient bolus $c_0$, the more mutations within a batch, leading to more species at the end of the batch (\emph{cf}. Fig.~\ref{fig:mutate}B). As $c_0/K$ increases, the number of extant species (species with non-zero abundance) increases since more growth events, and therefore more mutation events, occur within a batch. We also find that more evenly balanced nutrient supplies lead to a larger number of species. However, many of these species are very low abundance, and are recent mutations that will typically not survive more than a few batches. We therefore need to consider a metric which better reflects true diversity. 

A useful summary statistic for quantifying diversity \cite{Jost2006, Leinster2020} is the effective number of species $m_e = e^S$ with the Shannon diversity $S = -\sum_\sigma P_\sigma \ln P_\sigma$ and $P_\sigma = \rho^*_\sigma(0)/\rho_0$, with $\rho^*_\sigma(0)$ being the steady-state species abundances at the beginning of a batch. We show the effective number of species, $m_e$, as a function of $c_0$ in Fig.~\ref{fig:mutate}\emph{C}. As $c_0$ increases, the decrease in $m_e$ due to the early-bird effect and single-nutrient competition \cite{ErezLopez2020} is offset by mutations generating new species. As a result, for these parameters, $m_e$ is flat as $c_0 \approx K$. As $c_0$ increases further, $m_e$ does increase, due to both mutations and reduced remapping. This is evident in Fig.~\ref{fig:mutate}\emph{D} which shows more species and flatter rank-abundance curves for higher $c_0$ for a balanced nutrient supply (magenta). Even for an unbalanced nutrient supply (cyan), diversity increases for large enough $c_0/K$ (lower values of $c_0/K$ are shown in \emph{Appendix} Figs.~\ref{fig:mutations_withhigh_rhoK}, ~\ref{fig:mutations_withhigh_rhoK_nspecies}). 

\rr{This increase in diversity with increasing $c_0$ arises from a competition between the diversity-increasing effect of mutation and the diversity-reducing effect of demographic noise. Consider the growth dynamics within an individual batch. Mutations shift population to  non-dominant species, thus making the end-of-batch abundances on average more ``even'' as growth proceeds. However, this effect can be washed out by high levels of demographic noise, which can make abundances less equal. Demographic noise is high when the total population is small, so that each birth has a relatively large effect on the relative abundances. Thus, when $c_0$ and $\rho_0$ are small, demographic noise counteracts the diversity-increasing effects of mutation. As $c_0$ or $\rho_0$ increase, this demographic noise is reduced and therefore diversity rises.} 

\rr{To better understand this competition between mutation and demographic noise, we derived a Fokker-Planck equation for the dynamics of the probability distribution of the relative abundance of two species. For neutral growth with mutations, the population dynamics during a batch is given by }
	
\be
\frac{\partial P}{\partial \rho} =  \frac{\partial}{\partial x_1} \left( D \frac{\partial P }{\partial x_1}  \right) -\frac{\partial (P V )}{\partial x_1},
\label{eq:pde}
\ee	

\noindent \rr{where $P = P(x_1,\rho)$ is the probability distribution of the relative abundance $x_1$ of species 1 at total species abundance $\rho$, $D$ is the effective diffusion coefficient, and $V$ is the effective drift velocity. From the microscopic dynamics, we find that $D = \frac{(1-\nu)x_1(1-x_1)}{(\rho + 1)^2}$ and $V = \frac{\nu(1-2x_1)}{\rho + 1}$ (see \emph{Appendix} for details). }
	
\rr{The form of $D$ and $V$ reveal the contributions of demographic noise and mutation to the population dynamics. $D$ captures the effect of random births, it scales with $1 - \nu$ and drives the probability distribution towards the edges, being maximal at $x_1 = 0.5$. $V$ represents the effect of mutations: it scales with $\nu$ and drives the probability distribution towards the center, vanishing at $x_1 = 0.5$. The outcome of the competition between these two opposing effects is determined by the denominators of $D$ and $V$. Both $D$ and $V$ contain polynomials of the total population $\rho$ in their denominator, and therefore both effects weaken as growth proceeds during a batch (the larger the population, the smaller the effect of each birth on the relative abundance). However, the denominator of $D$ is quadratic in $\rho$, while the denominator of $V$ is only linear in $\rho$. As a result, the relative strength of mutation increases as growth proceeds, driving the system towards a $\delta$ function at $x_1 = 0.5$ as the population becomes very large. To demonstrate this, we show a numerical simulation of Eq. \ref{eq:pde} in Fig. \ref*{fig:mutate}E. Beginning with a narrow distribution of abundances in a small population, $P$ rapidly widens once growth begins due to demographic noise. As growth proceeds, the distribution becomes narrow once again as demographic noise decreases in strength relative to the equalizing effect of mutation. }

Broadly speaking, mutations in our model lead to a ``rich-get-poorer'' effect in which high-abundance species feed low-abundance species with a steady stream of mutants, countering the ``rich-get-richer'' impact of the early-bird effect and competition for a single dominant nutrient. \rr{However, because growth in this system is stochastic, for small enough initial populations this ``rich-get-poorer'' effect must first overcome the diversity-reducing effects of demographic noise.}

\section*{Discussion}
In nature, microbial metabolic strategies vary within a single generation and across generations. In this work, we have built on existing models of resource competition under serial dilution by exploring what happens when species are able to modify their metabolisms. We considered two types of strategy change: transient regulatory changes and random mutations. Interestingly, these two mechanisms have substantial but drastically different effects on community diversity, highlighting the potential impact of changes in metabolic strategy on real microbial communities. 

We first considered the outcome of introducing an ``adapter", a species capable of regulating its enzyme allocation, into a stable diverse ecosystem. We found that, over long time periods, the adapter curtailed diversity in a manner similar to introducing a species with an enhanced enzyme budget \cite{Posfai2017}. This can be viewed as an augmented early-bird effect. In our previous model, an early-bird can take advantage of fast initial growth to rapidly consume all nutrients, even those it consumes inefficiently. An adapter can tune its enzyme levels to first efficiently consume the more valuable nutrient, and then switch enzyme production to focus on the remaining nutrients. Thus, it benefits from both an early population advantage and opportune enzyme allocation. As a result, the invading adapter gains a small but significant fitness advantage and gradually takes over the community. In some cases, the rest of the community is able to self-organize such that the differences in nutrient concentrations became too small for the adapter to detect, precluding its advantage. Intriguingly, in some cases, the adapter only reaches dominance over a very long, emergent timescale. In that case, fluctuations in a real ecosystem might wash away the adapter's advantage. In summary, the impact of enzyme regulation by some species in an ecosystem depends on both the metabolic cost of maintaining enzyme regulation and the presence of other timescales in the system. Studying the relevance of the long timescale introduced by enzyme regulation in shaping ecosystem diversity is a promising direction of future study. 

\rr{In this work, we have focused on competition between adapters and non-adapters, but we did not consider the scenario of competition between adapters. On evolutionary timescales, this scenario is likely to arise as the adapters themselves will mutate and speciate. This is a worthwhile direction of future study, as it is not clear that adaptation will collapse diversity in the context of adapter-adapter competition. Interestingly, a recently published work analyzing a conceptually similar model of enzyme regulation showed that competition between adapters can stabilize diversity \cite{PaccianiMori2020}. While this model differs from the one we analyze here in the regulation scheme employed, these results suggest the intriguing possibility that adaptation can collapse diversity when it first arises, but promote diversity once the adapters themselves speciate.}

In contrast to our results on adaptation, we found that the addition of \rr{mutations that randomize metabolic strategy} modified the relationship between bolus size and diversity to be monotonically increasing. This is a dramatic change from the model without mutation, where the relationship is generally non-monotonic \cite{ErezLopez2020}. However, as with the original model, this behavior can be understood in terms of the early-bird effect. Without mutation, the early-bird effect leads to a ``rich-get-richer" effect that initially decreases diversity as bolus size increases. This occurs because increasing the supply of nutrients leads to the additional nutrients being disproportionately taken up by the most abundant species. Indeed, until nutrients become saturating in the $c_0 \gg K$ limit, the more nutrient, the less diversity. However, the addition of mutation opposes this one-sided concentration of biomass, acting somewhat similarly to ``income tax'': the species that consume the most nutrients and therefore proliferate fastest, are the species that lose the largest fraction of their population each batch to mutations. As the bolus size grows, the number of birth events (and therefore mutation events) increases, thereby increasing overall diversity by \rr{redistributing a larger fraction of the total population from more abundant to less abundant species}. \rr{Our simulations have focused on a particular class of strategy-randomizing mutations, and so, it would be interesting to see if our observations generalize beyond this particular choice of mutational effect}.

Our exploration of the early-bird effect and the adapter provides some insight into the enzyme regulation strategies utilized by real microbes. When supplied with high concentrations of nutrients (corresponding in our model to a large nutrient bolus) real microbes are known to utilize a diauxic strategy in which they will exclusively consume the most valuable nutrient until is entirely depleted before switching to metabolism of less valuable nutrients \cite{Monod1942}. This strategy is entirely consistent with the optimal enzyme regulation needed to exploit the early-bird effect. It is better to devote all resources to the nutrient that allows for the highest growth, and then use the early-bird advantage to more efficiently exploit the remaining nutrients. Interestingly, it has been found that in environments containing low nutrient levels (corresponding to the small bolus size limit of our model), microbes instead employ a mixed-utilization strategy where they attempt to consume multiple different types of nutrients \cite{Egli1993,KovarovaKovar1998}. This is also consistent with our model, in the low nutrient limit the early-bird effect is weak or non-existent, lowering the benefit of sophisticated regulatory strategies (though growth on multiple nutrients simultaneously could also arise if individual nutrient concentrations are too low to support growth on that nutrient alone). These results highlight the fact that certain effects cannot be found in chemostat models, and therefore models with fluctuating nutrient supply have an important role in efforts to understand microbial ecosystems.

In our previous work, we argued that in order to understand microbial diversity, it is necessary to take into account fluctuations in the environment. Here, we have shown that variations within the microbes themselves can also play a key role in microbial diversity. Indeed, these two types of fluctuations can interact in a complex manner, as is the case with adapter's exploitation of the early-bird effect. On a practical level, our results suggest that measuring microbial abundances and environmental conditions may not be sufficient to understand diversity in microbial ecology experiments. Even in a simple model, if some microbes can adapt to ambient conditions, they may shape the ecosystem in complicated ways over very long timescales. And so, it appears that predictive models of microbial ecosystem dynamics would benefit from information about the microbes' inner states and decision-making processes. 
 
\emph{Acknowledgements.} This work was supported by National Institutes of Health (www.nih.gov) Grant R01 GM082938 (N.S.W.), and National Science Foundation Grant GRFP DGE-1656466 (J.G.L.) and through the Center for the Physics of Biological Function (PHY-1734030). This research was supported in part by NSF Grant No. PHY-1748958, NIH Grant No. R25GM067110, and the Gordon and Betty Moore Foundation Grant No. 2919.01. 

N.S.W. supervised the research; A.E. and J.G.L. wrote simulations and did analytic calculations; all authors interpreted results; A.E., J.G.L., and N.S.W. wrote the paper. 

%All code and data used in this manuscript available at \href{https://github.com/AmirErez/SeasonalEcosystem}{https://github.com/AmirErez/SeasonalEcosystem}.

\bibliographystyle{unsrt}
%\bibliography{AdaptationMutation}

\begin{thebibliography}{}
\bibitem{Bienhold2012}
Christina Bienhold, Antje Boetius, and Alban Ramette.
\newblock The energy-diversity relationship of complex bacterial communities in
  arctic deep-sea sediments.
\newblock {\em The ISME Journal}, 6:724, November 2012.

\bibitem{Smits2017}
Samuel~A. Smits, Jeff Leach, Erica~D. Sonnenburg, Carlos~G. Gonzalez, Joshua~S.
  Lichtman, Gregor Reid, Rob Knight, Alphaxard Manjurano, John Changalucha,
  Joshua~E. Elias, Maria~Gloria Dominguez-Bello, and Justin~L. Sonnenburg.
\newblock Seasonal cycling in the gut microbiome of the hadza hunter-gatherers
  of tanzania.
\newblock {\em Science}, 357(6353):802--806, 2017.

\bibitem{gajer2012temporal}
Pawel Gajer, Rebecca~M Brotman, Guoyun Bai, Joyce Sakamoto, Ursel~ME
  Sch{\"u}tte, Xue Zhong, Sara~SK Koenig, Li~Fu, Zhanshan~Sam Ma, Xia Zhou,
  et~al.
\newblock Temporal dynamics of the human vaginal microbiota.
\newblock {\em Science translational medicine}, 4(132):132ra52--132ra52, 2012.

\bibitem{Lloyd-Price2016}
Jason Lloyd-Price, Galeb Abu-Ali, and Curtis Huttenhower.
\newblock The healthy human microbiome.
\newblock {\em Genome Medicine}, 8(1):51, Apr 2016.

\bibitem{Ladau2013}
Joshua Ladau, Thomas~J Sharpton, Mariel~M Finucane, Guillaume Jospin, Steven~W
  Kembel, James O'dwyer, Alexander~F Koeppel, Jessica~L Green, and Katherine~S
  Pollard.
\newblock Global marine bacterial diversity peaks at high latitudes in winter.
\newblock {\em The ISME journal}, 7(9):1669--1677, 2013.

\bibitem{Weigel2019}
Brooke~L. Weigel and Catherine~A. Pfister.
\newblock Successional dynamics and seascape-level patterns of microbial
  communities on the canopy-forming kelps nereocystis luetkeana and macrocystis
  pyrifera.
\newblock {\em Frontiers in Microbiology}, 10:346, 2019.

\bibitem{Ptacnik2008}
Robert Ptacnik, Angelo~G Solimini, Tom Andersen, Timo Tamminen, P{\aa}l
  Brettum, L.~Lepisto, E.~Willen, and Seppo Rekolainen.
\newblock {Diversity predicts stability and resource use efficiency in natural
  phytoplankton communities}.
\newblock {\em Proc. Natl. Acad. Sci. U.S.A.}, 105(13):5134--5138, 2008.

\bibitem{VanElsas2012}
J.~D. van Elsas, M.~Chiurazzi, C.~A. Mallon, D.~Elhottova, V.~Kristufek, and
  J.~F. Salles.
\newblock {Microbial diversity determines the invasion of soil by a bacterial
  pathogen}.
\newblock {\em Proc. Natl. Acad. Sci. U.S.A.}, 109(4):1159--1164, 2012.

\bibitem{Taur2014}
Ying Taur, Robert~R. Jenq, Miguel~Angel Perales, Eric~R. Littmann, Sejal
  Morjaria, Lilan Ling, Daniel No, Asia Gobourne, Agnes Viale, Parastoo~B.
  Dahi, Doris~M. Ponce, Juliet~N. Barker, Sergio Giralt, Marcel {Van Den
  Brink}, and Eric~G. Pamer.
\newblock {The effects of intestinal tract bacterial diversity on mortality
  following allogeneic hematopoietic stem cell transplantation}.
\newblock {\em Blood}, 124(7):1174--1182, 2014.

\bibitem{Stein2013}
Richard~R. Stein, Vanni Bucci, Nora~C. Toussaint, Charlie~G. Buffie, Gunnar
  R{\"a}tsch, Eric~G. Pamer, Chris Sander, and Jo{\~a}o~B. Xavier.
\newblock Ecological modeling from time-series inference: Insight into dynamics
  and stability of intestinal microbiota.
\newblock {\em PLoS Computational Biology}, 9(12):e1003388, December 2013.

\bibitem{Levin1970}
Simon~A Levin.
\newblock {Community Equilibria and Stability, and an Extension of the
  Competitive Exclusion Principle}.
\newblock {\em Am. Nat.}, 104(939):413--423, 1970.

\bibitem{Armstrong1980}
Robert~A Armstrong and Richard McGehee.
\newblock {Competitive Exclusion}.
\newblock {\em Am. Nat.}, 115(2):151--170, 1980.

\bibitem{Hutchinson1961}
G.~E. Hutchinson.
\newblock The paradox of the plankton.
\newblock {\em Am. Nat.}, 95(882):137--145, May 1961.

\bibitem{Goyal2018}
Akshit Goyal and Sergei Maslov.
\newblock {Diversity, Stability, and Reproducibility in Stochastically
  Assembled Microbial Ecosystems}.
\newblock {\em Physical Review Letters}, 120(15):158102, 2018.

\bibitem{Kelsic2015}
Eric~D. Kelsic, Jeffrey Zhao, Kalin Vetsigian, and Roy Kishony.
\newblock {Counteraction of antibiotic production and degradation stabilizes
  microbial communities}.
\newblock {\em Nature}, 521(7553):516--519, 2015.

\bibitem{Thingstad2000}
Tron~Frede Thingstad.
\newblock {Elements of a theory for the mechanisms controlling abundance,
  diversity, and biogeochemical role of lytic bacterial viruses in aquatic
  systems}.
\newblock {\em Limnology and Oceanography}, 45(6):1320--1328, 2000.

\bibitem{PhysRevLett.119.268101}
Chi Xue and Nigel Goldenfeld.
\newblock Coevolution maintains diversity in the stochastic ``kill the winner''
  model.
\newblock {\em Phys. Rev. Lett.}, 119:268101, Dec 2017.

\bibitem{Murrell2003}
David~J. Murrell and Richard Law.
\newblock Heteromyopia and the spatial coexistence of similar competitors.
\newblock {\em Ecology Letters}, 6(1):48--59, 2003.

\bibitem{Tilman1994}
David Tilman.
\newblock Competition and biodiversity in spatially structured habitats.
\newblock {\em Ecology}, 75(1):2--16, 1994.

\bibitem{scheffer2003plankton}
Marten Scheffer, Sergio Rinaldi, Jef Huisman, and Franz~J Weissing.
\newblock Why plankton communities have no equilibrium: solutions to the
  paradox.
\newblock {\em Hydrobiologia}, 491(1):9--18, 2003.

\bibitem{Posfai2017}
Anna Posfai, Thibaud Taillefumier, and Ned~S. Wingreen.
\newblock Metabolic trade-offs promote diversity in a model ecosystem.
\newblock {\em Phys. Rev. Lett.}, 118:028103, Jan 2017.

\bibitem{Weiner2019}
Benjamin~G. Weiner, Anna Posfai, and Ned~S. Wingreen.
\newblock Spatial ecology of territorial populations.
\newblock {\em Proceedings of the National Academy of Sciences}, 116(36):17874,
  09 2019.

\bibitem{ErezLopez2020}
Amir Erez, Jaime~G Lopez, Benjamin~G Weiner, Yigal Meir, and Ned~S Wingreen.
\newblock Nutrient levels and trade-offs control diversity in a serial dilution
  ecosystem.
\newblock {\em Elife}, 9:e57790, 2020.

\bibitem{li2020modeling}
Zhiyuan Li, Bo~Liu, Sophia Hsin-Jung Li, Christopher~G King, Zemer Gitai, and
  Ned~S Wingreen.
\newblock Modeling microbial metabolic trade-offs in a chemostat.
\newblock {\em PLoS computational biology}, 16(8):e1008156, 2020.

\bibitem{PaccianiMori2020}
Leonardo Pacciani-Mori, Andrea Giometto, Samir Suweis, and Amos Maritan.
\newblock Dynamic metabolic adaptation can promote species coexistence in
  competitive microbial communities.
\newblock {\em PLOS Computational Biology}, 16(5):e1007896, 2021/04/07/05:50:02
  2020.

\bibitem{Palmer1994}
Michael~W. Palmer.
\newblock {Variation in species richness: Towards a unification of hypotheses}.
\newblock {\em Folia Geobotanica et Phytotaxonomica}, 29(4):511--530, 1994.

\bibitem{Lenski}
RE~Lenski and M~Travisano.
\newblock {Dynamics of Adaptation and Diversification}.
\newblock {\em Proc. Natl. Acad. Sci. U.S.A.}, 91(July):6808--6814, 1994.

\bibitem{Goldford2018}
Joshua~E. Goldford, Nanxi Lu, Djordje Baji{\'c}, Sylvie Estrela, Mikhail
  Tikhonov, Alicia Sanchez-Gorostiaga, Daniel Segr{\`e}, Pankaj Mehta, and
  Alvaro Sanchez.
\newblock Emergent simplicity in microbial community assembly.
\newblock {\em Science}, 361(6401):469, August 2018.

\bibitem{shimizu2014regulation}
Kazuyuki Shimizu.
\newblock Regulation systems of bacteria such as escherichia coli in response
  to nutrient limitation and environmental stresses.
\newblock {\em Metabolites}, 4(1):1--35, 2014.

\bibitem{bajic2020ecology}
Djordje Bajic and Alvaro Sanchez.
\newblock The ecology and evolution of microbial metabolic strategies.
\newblock {\em Current opinion in biotechnology}, 62:123--128, 2020.

\bibitem{Good2018}
Benjamin~H. Good, Stephen Martis, and Oskar Hallatschek.
\newblock Adaptation limits ecological diversification and promotes ecological
  tinkering during the competition for substitutable resources.
\newblock {\em Proc Natl Acad Sci USA}, 115(44):E10407, October 2018.

\bibitem{ramiro2020low}
Ricardo~S Ramiro, Paulo Dur{\~a}o, Claudia Bank, and Isabel Gordo.
\newblock Low mutational load and high mutation rate variation in gut commensal
  bacteria.
\newblock {\em PLoS biology}, 18(3):e3000617, 2020.

\bibitem{friedman2013sympatric}
Jonathan Friedman, Eric~J Alm, and B~Jesse Shapiro.
\newblock Sympatric speciation: when is it possible in bacteria?
\newblock {\em PLoS One}, 8(1):e53539, 2013.

\bibitem{Bialek10040}
William Bialek and Sima Setayeshgar.
\newblock Physical limits to biochemical signaling.
\newblock {\em Proceedings of the National Academy of Sciences},
  102(29):10040--10045, 2005.

\bibitem{Jost2006}
Lou Jost.
\newblock Entropy and diversity.
\newblock {\em Oikos}, 113(2):363--375, May 2006.

\bibitem{Leinster2020}
Tom Leinster.
\newblock Entropy and diversity: The axiomatic approach, 2020.

\bibitem{Monod1942}
Jacques Monod.
\newblock {\em Recherches sur la croissance des cultures bacte{\'e}riennes}.
\newblock Hermann and Cie, 1942.

\bibitem{Egli1993}
Thomas Egli, Urs Lendenmann, and Mario Snozzi.
\newblock Kinetics of microbial growth with mixtures of carbon sources.
\newblock {\em Antonie van Leeuwenhoek}, 63(3):289--298, Sep 1993.

\bibitem{KovarovaKovar1998}
Karin Kov{\'a}rov{\'a}-Kovar and Thomas Egli.
\newblock Growth kinetics of suspended microbial cells: From
  single-substrate-controlled growth to mixed-substrate kinetics.
\newblock {\em Microbiology and Molecular Biology Reviews}, 62(3):646--666,
  1998.
 

\end{thebibliography}

\newpage
\clearpage

\appendix

\section*{Appendix}
\label{sec:methods}
This section describes the simulation methods used in this manuscript. All code and data used in this manuscript can be found at \href{https://github.com/AmirErez/SeasonalEcosystem}{https://github.com/AmirErez/SeasonalEcosystem}.

\subsubsection*{Deterministic dynamics}

We numerically solve the ODEs within each batch using a custom MATLAB-coded fourth-order Runge-Kutta solver with adaptive step size. Step size at a given time step is chosen such that the relative change of all state variables is below a predetermined threshold. 

\rr{To numerically solve the Fokker-Planck model of mutation, we utilize a custom MATLAB-coded first-order Euler solver. The interval $[0,1]$ was discretized into 801 points and all derivatives were computed using first-order centered differences. Zero-flux conditions were imposed at the boundaries of the domain. }

\subsubsection*{Population bottleneck sampling}

We implement discrete sampling when diluting from one batch to the next by picking without replacement $\rho_0$ individuals from a total end-of-batch population of $\rho_0+c_0$. If there are non-integer populations at the end of a batch (as can occur with deterministic dynamics), they are rounded up if $\rho_\sigma -\mbox{floor}(\rho_\sigma) > U(0,1)$ where floor rounds down to the nearest integer and $U(0,1)$ is a uniform random variable between 0 and 1. For all simulations with stochastic bottlenecks, we allow the simulation to equilibrate for 10,000 dilutions and average over 90,000 further dilutions.

\hspace{4mm}

\subsubsection*{Mutation-selection dynamics}

We use Gillespie's algorithm to simulate the reactions shown in Table ~\ref{tab:Gillepsie} until all nutrients $\{c_i\}$ are depleted.

\begin{table}[!h]
	\centering	
	\begin{tabular}{lll}
		\hline
		\emph{Name} & \emph{Reaction} & \emph{Rate} \\
		\hline
		Birth & $\sigma \rightarrow 2\sigma$ & $(1-\nu)\rho_\sigma \sum_i j_{\sigma,i}$ \\
		Mutation & $\sigma \rightarrow \sigma + \sigma'$ & $\nu \rho_\sigma \sum_i j_{\sigma,i}$ with randomly chosen $\sigma'$ \\
		Time & $t\rightarrow t + \Delta t$ & $\Delta t = -\ln(U(0,1))/ \sum_{\sigma,i} \rho_\sigma j_{\sigma,i} $ \\
		\hline
	\end{tabular} 
	\caption{Gillespie reactions for mutation-selection dynamics}
	\label{tab:Gillepsie}
\end{table}
\noindent with $U(0,1)$ a uniform random variable between 0 and 1. For each ``birth'' reaction, ($\rho_\sigma$ increases by 1), $c_i$ decreases by $j_{\sigma,i} / \sum_i{j_{\sigma,i}}$.

\bigskip
\noindent For all simulations featuring mutation (Fig.~\ref{fig:mutate} and \emph{Appendix} Figs.~\ref{fig:DilutionRegimes}-\ref{fig:mutations_withhigh_rhoK_nspecies}), we let the system equilibrate over 10,000 dilutions, many more than required to reach steady state in the deterministic model, and then average over at least 90,000 more dilutions. To ensure that the results do not depend on the number of possible strategies, (due to mutations saturating all possible species), we increase the total number to 201 species equally spaced between 0 and 1.

\newpage
\clearpage
\twocolumngrid

\subsubsection*{Derivation of Fokker-Planck equation for mutation model}
\label{sec:pde}
\rr{We consider a batch culture containing two species with equal fitness. The dynamics of the growth process can be described as a random walk of the relative abundance $x_1$ of species 1 on the interval $[0,1]$. The dynamics of the probability distribution of $x_1$ as a function of total abundance in the batch $\rho$ can be approximated as a Fokker-Planck equation with effective diffusion coefficient $D$ and effective drift velocity $V$:}

\be
\frac{\partial P}{\partial \rho} =  \frac{\partial}{\partial x_1} \left( D \frac{\partial P }{\partial x_1}  \right) -\frac{\partial (P V )}{\partial x_1}.
\ee	

\noindent \rr{The form of $V$ and $D$ can be computed from the microscopic dynamics. Consider a batch containing $\rho$ cells, $n$ of which belong to species 1. The drift velocity, $V$, will be the expected value of the change in relative abundance from mutation events:}

\be
\begin{split}
V(x_1,\rho) = \nu \left( \frac{n}{\rho} \right) \left( \frac{n}{\rho + 1} - \frac{n}{\rho} \right) + \\ \nu \left( \frac{\rho - n}{\rho} \right) \left( \frac{n+1}{\rho + 1} - \frac{n}{\rho} \right).
\end{split}
\ee

\noindent \rr{This expression reduces to }

\be
V(x_1,\rho) = \frac{\nu (1-2x_1)}{\rho + 1}.
\ee

\noindent \rr{Similarly, the diffusion coefficient, $D$, will be the variance of the change in relative abundance due to stochastic neutral growth}

\be
\begin{split}
	V(x_1,\rho) = (1-\nu) \left( \frac{\rho-n}{\rho} \right) \left( \frac{n}{\rho + 1} - \frac{n}{\rho} \right)^2 + \\ (1-\nu) \left( \frac{n}{\rho} \right) \left( \frac{n+1}{\rho + 1} - \frac{n}{\rho} \right)^2,
\end{split}
\ee

\noindent \rr{which simplifies to}

\be
D(x_1,\rho) = \frac{(1-\nu)x_1(1-x_1)}{(\rho+1)^2}.
\ee

\pagebreak
\newpage

\subsubsection*{Appendix Figures} 
%\begin{figure}
%\centering
%\includegraphics[width=0.25\textwidth]{Fig_stochremap.eps}
%%\includegraphics[width=0.5\columnwidth]{Fig_stochremap.eps}
%\caption{Remapping under stochastic Gillespie dynamics without mutation ($\nu=0$). With stochastic growth dynamics, the remapping itself becomes stochastic with a large variance for $c_0/K \ll 1$. For every observation of remapping at a given strategy and $c_0/K$ value, the growth-function integrals $I_i$ were computed across the range of nutrient supply proportions. Details about $I_i$ could be found in Ref~\cite{ErezLopez2020}. The remapped coexistence boundary for that set of simulations is the nutrient supply which produced the smallest difference between $I_1$ and $I_2$. For a given strategy and $c_0/K$ value, this procedure was repeated multiple times to obtain a remapping density. Here, the log of the remapping density is shown as a heatmap for $\rho_0/K=10^{-3}$. Colored curves show the corresponding deterministic remapping of 21 equally spaced strategies, as in Fig.~3\emph{A} from Ref~\cite{ErezLopez2020}.}
%\label{fig:stoch_remapping}
%\end{figure}

\begin{figure}[!htbp]
	\includegraphics[width=0.49\columnwidth]{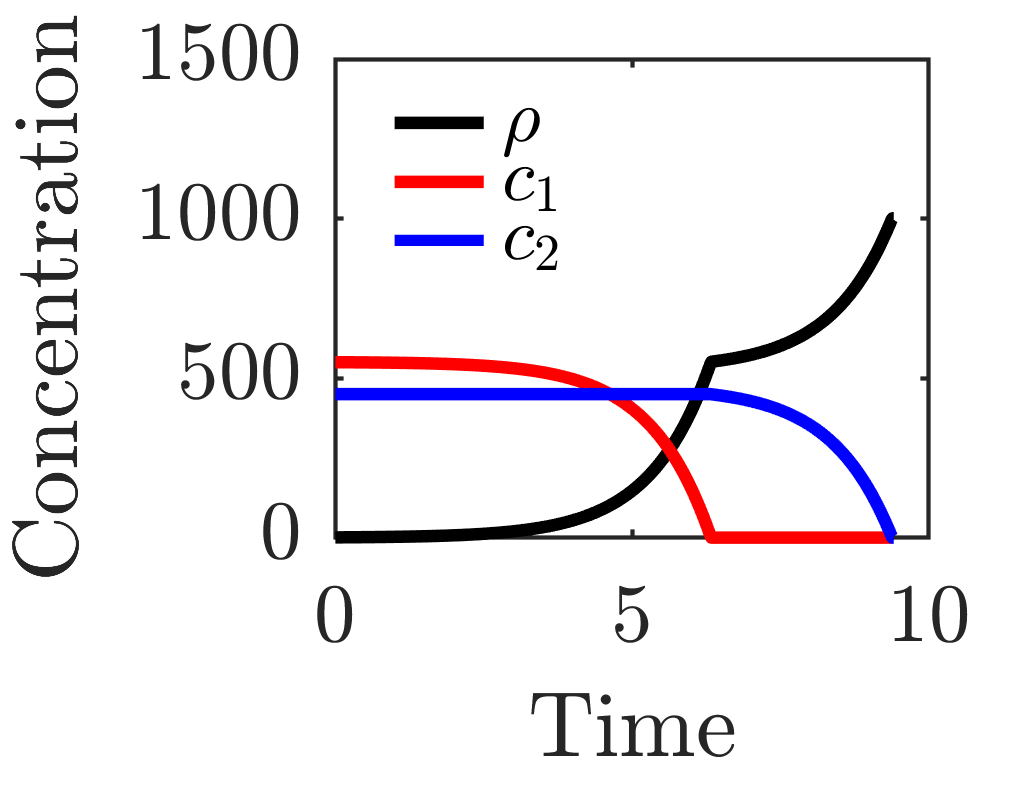}
	\caption{\rr{Example of diauxie in the adaptation model. When sensing tolerances are large, the adaptation model can exhibit diauxie, in which nutrients are consumed sequentially rather than simultaneously. In the simulation shown, $\rho_0 = K = 1$, $c_0 = 10^3$, $\Delta_c = 0.95$, and Nutrient 1 fraction $= 0.55$.}}
	\label{fig:diauxie_example}
\end{figure}

\begin{figure}[!htbp]
\includegraphics[width=0.49\columnwidth]{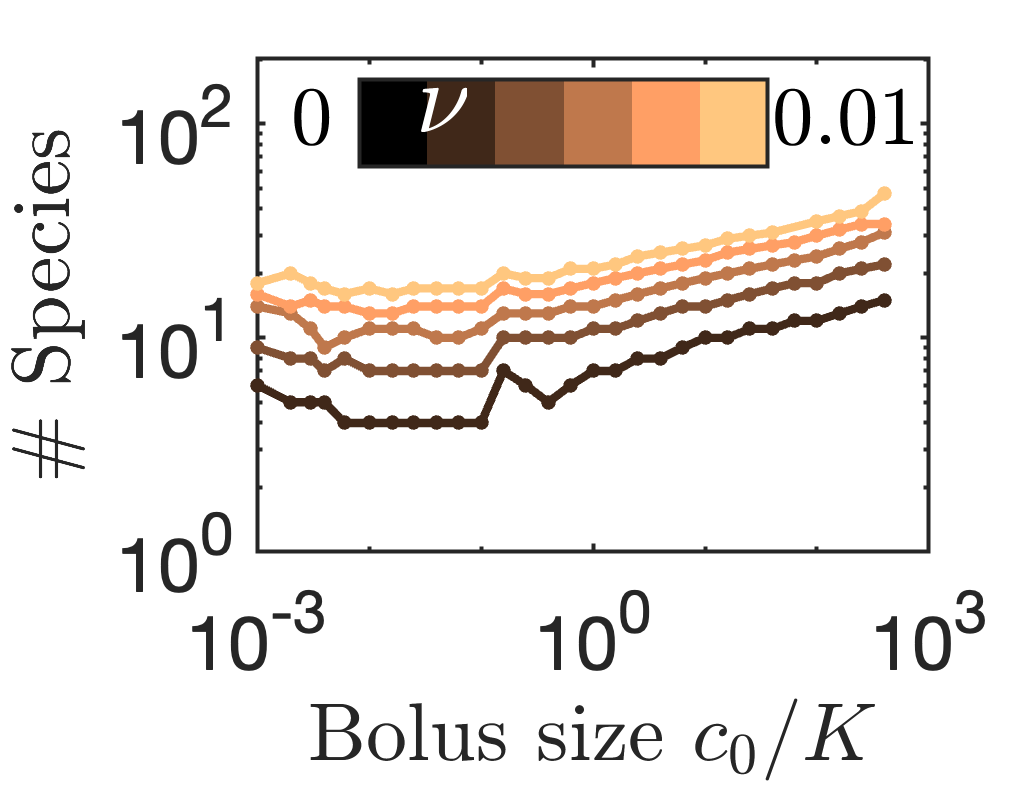}
\caption{Median number of extant species as a function of $c_0/K$ at different mutation rates $\nu$ from 0.002 (dark) to 0.01 (orange) with Nutrient 1 fraction 0.005. With increasing $\nu$, more species are created by mutation during each batch. When $\nu=0$, the median number of extant species fluctuates between one and two species due to extinctions though sampling noise. The steady state reflects a balance between the addition of new species through mutation and the loss of species due to inter-batch sampling.}
\label{fig:nspecies_in_maintext}
\end{figure}

\smallskip

\begin{figure}[!htbp]
\centering
\includegraphics[width=0.98\columnwidth]{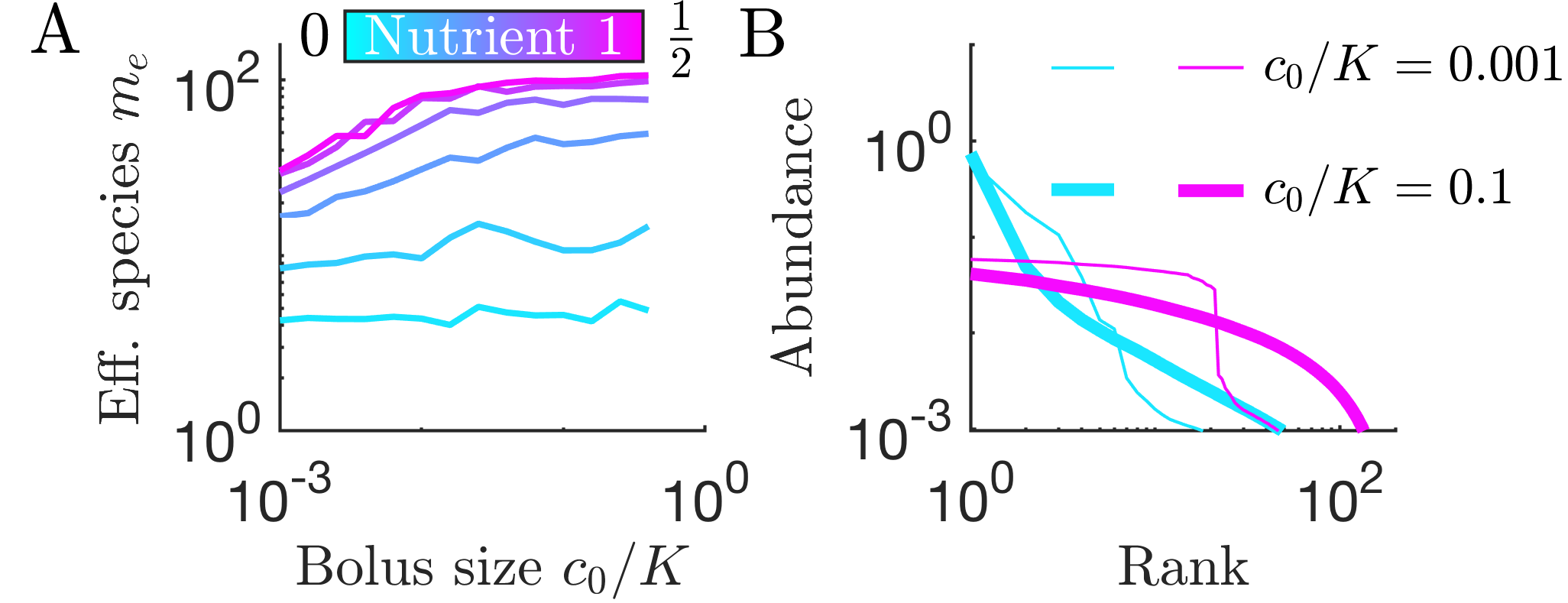}
\caption{Mutation-selection balance for high $\rho_0$ and $K$ ($\rho_0=10^5$, $K=10^5$, and $\nu=10^{-3}$). The results presented in Fig.~\ref{fig:mutate} in the main text are for $\rho_0=1000$, $K=1000$, and $\nu=0.01$; with those parameters, sampling noise dominates in the $c_0/K \ll 1$ limit. Here, we explore the $c_0/K \ll 1$ limit using larger $\rho_0$ and $K$. (\emph{A}) Effective number of species $m_e$ for different nutrient compositions (colors) as a function of nutrient bolus size $c_0/K$. (\emph{B}) Rank-abundance curves for Nutrient 1 fraction 0.05 (cyan) and 0.5 (magenta); line thickness corresponds to $c_0/K$ values. We note the similar trends to the results in the main text.}
\label{fig:mutations_withhigh_rhoK}
\end{figure}

\begin{figure}[!htbp]
\centering
\includegraphics[width=0.49\columnwidth]{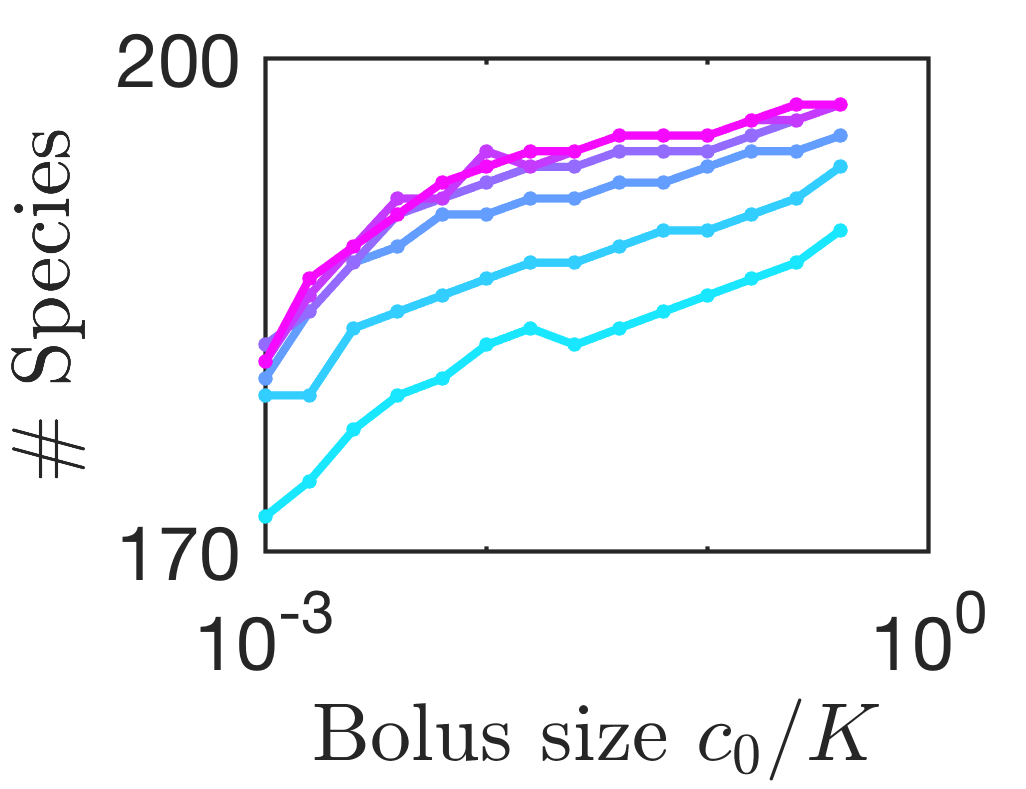}
\caption{Number of extant species under mutation-selection balance for $\rho_0=10^5$, $K=10^5$, and $\nu=10^{-3}$. Note the monotonic increase in the median number of extant species with $c_0$ without saturating the total number of possible species (here 201).}
\label{fig:mutations_withhigh_rhoK_nspecies}
\end{figure}

\newpage

\begin{figure*}[!htbp]
	\includegraphics[width=0.33\textwidth]{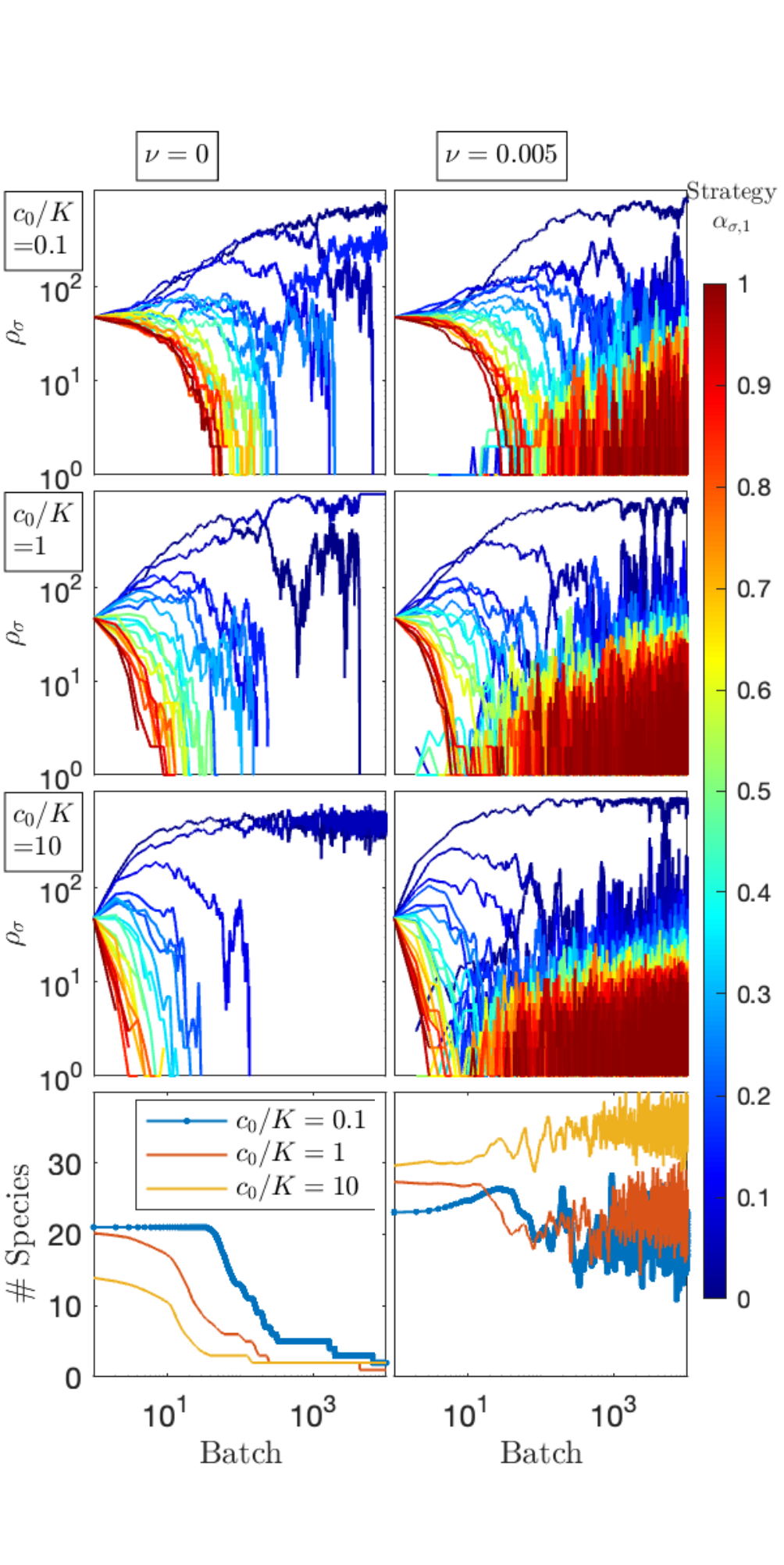}
	\includegraphics[width=0.33\textwidth]{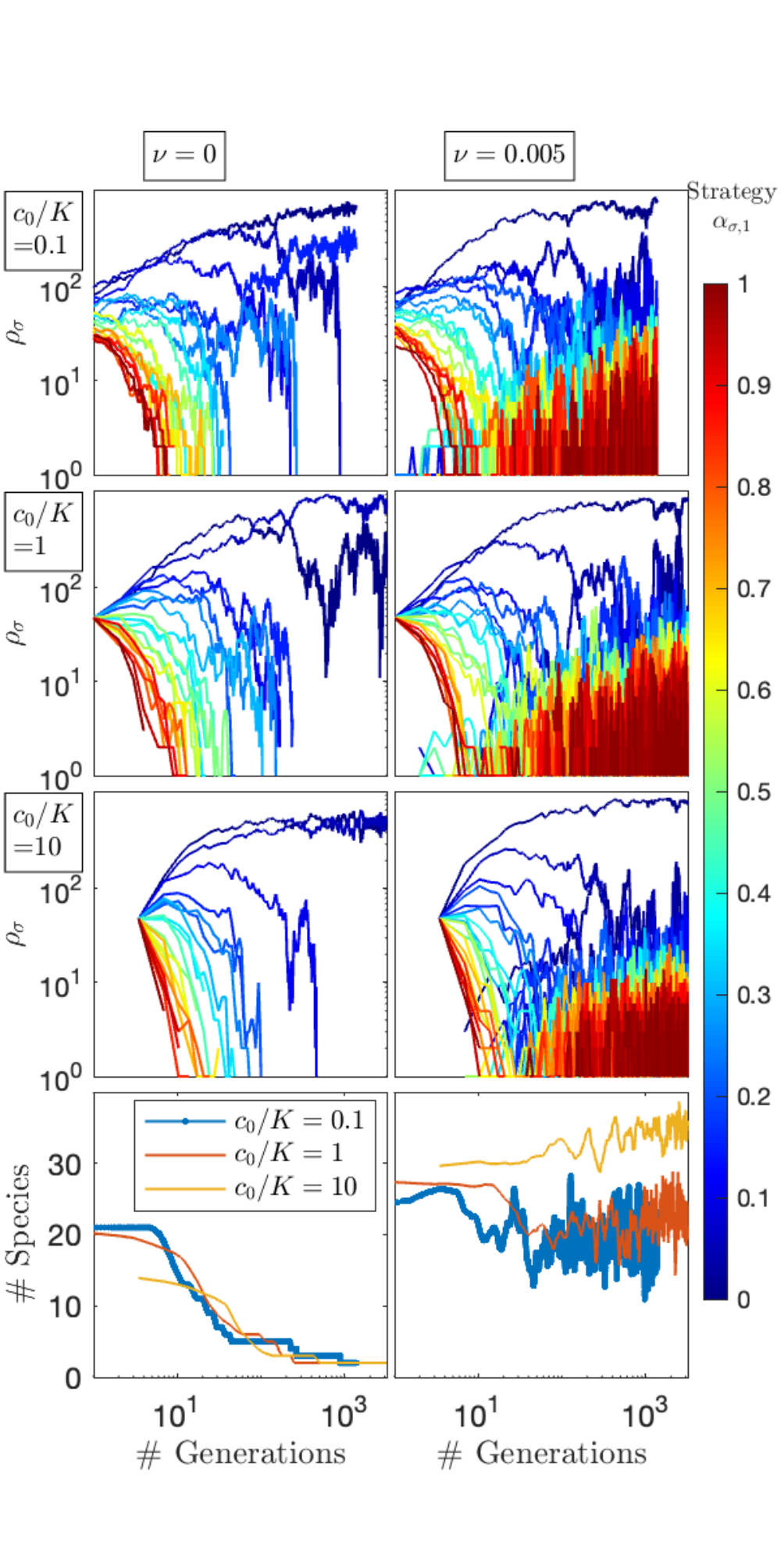}
	\caption{\textbf{Left two columns:} Timecourses of two stochastic models: without mutation ($\nu=0$) and with mutation ($\nu=0.005$). In the main text, we only present data on the steady states of the stochastic models while here we examine the full timecourses. In all cases, inter-batch sampling is stochastic without replacement. In the model without mutations ($\nu=0$), sampling causes extinctions of species with no possibility of recovery. The model with mutations results in a fluctuating number of species, as species constantly go extinct and are reborn through mutation. $x$-axis is batch number. \textbf{Right two columns:} Timecourses of two stochastic models, as the left two columns, but rescaling the $x$-axis to \rr{count the number of generations. The fractional population increase in a single batch with $c_0$ nutrients is $\frac{\rho_0+c_0}{\rho_0}$ and so, the number of elapsed generations is $\mbox{\# batches}\times \log_2 \left(\frac{\rho_0+c_0}{\rho_0} \right)$.}}
	\label{fig:DilutionRegimes}
\end{figure*}

\smallskip

%\begin{figure}
%	\includegraphics[width=0.49\textwidth]{SupplDilutionRegimesRenormx.eps}
%	\caption{Timecourses of three stochastic models, as in Fig.~\ref{fig:DilutionRegimes}, but rescaling the $x$-axis to approximate true time. We scale the number of batches by the time it takes to consume $c_0$ nutrients as $\log \left( \frac{\rho_0+c_0}{\rho_0}\right)$. This approximation comes from assuming saturated growth dynamics ($\rho(t) = \rho_0 e^{Et}$) and computing the time it takes to consume $c_0$ nutrients.} 
%	\label{fig:DilutionRegimesRenormx}
%\end{figure}

\end{document}